\documentstyle[prl,twocolumn,aps]{revtex}
\begin{document}
\hyphenation{tem-per-at-ure}
\draft
\wideabs{
\title{CuSiO$_3$ : a quasi - one - dimensional $S=1/2$ \\
antiferromagnetic chain system}
\author{M. BAENITZ, C. GEIBEL, M. DISCHNER, G. SPARN, F. STEGLICH }
\address{Max-Planck-Institute of Chemical Physics of Solids, D-01187 Dresden,
Germany}
\author{H. H. OTTO, M. MEIBOHM}
\address{TU - Clausthal, Institut f\"ur Mineralogie und Mineralische Rohstoffe,\\ Adolph - Roemer -Stra{\ss}e 2a, 38678 Clausthal - Zellerfeld,
Germany}
\author{A. A. GIPPIUS}
\address{Faculty of Physics, Moscow State University, 119899 Moscow, Russia}
\date{17.05.2000}
\maketitle
\begin{abstract}
{\ CuSiO$_3$ , isotypic to the spin - Peierls compound CuGeO$_3$,
was discovered recently as a metastable decomposition product of
the silicate mineral dioptase, Cu$_6$Si$_6$O$_{18}\cdot$6H$_2$O.
We investigated the physical properties of CuSiO$_3$ using
susceptibility,  magnetization and specific heat measurements on
powder samples. The magnetic susceptibility $\chi(T)$ is
reproduced very well above T = 8 K by theoretical calculations for
an S=1/2 antiferromagnetic Heisenberg linear chain without
frustration ($\alpha = 0$) and a nearest - neighbor exchange
coupling constant of $J/k_{B} = 21$\, K, much weaker than in
CuGeO$_3$. Below $8$\, K the susceptibility exhibits a substantial
drop. This feature is identified as a second - order phase
transition at $T_{0} = 7.9$\, K by specific heat measurements. The
influence of magnetic fields on $T_{0}$ is weak, and ac -
magnetization measurements give strong evidence for a spin - flop
- phase at $\mu_0H_{SF}$ $\simeq 3$\, T. The origin of the
magnetic phase transition at $T_{0} = 7.9$\, K is discussed in the
context of long - range antiferromagnetic order (AF) versus spin -
Peierls(SP)order. Susceptibility and specific heat results support
the AF ordered ground state. Additional temperature dependent
$^{63,65}Cu$ nuclear quadrupole resonance experiments have been
carried out to probe the $Cu^{2+}$ electronic state and the spin
dynamics in CuSiO$_3$.}
\end{abstract}

\pacs{Pacs: 75.40.Cx; 75.30.Et; 75.10.Jm} }
\section{Introduction}
The one dimensional spin system CuGeO$_3$ has attracted
considerable attention in the past years, since it was the first
(and up to now the only one) inorganic compound which exhibits a
spin - Peierls transition (SP). \cite{Hase93,Boucher96} The
simplicity of crystal growth and the large variety of possible
substitutions on the Cu and Ge sites promoted a huge amount of
experimental and theoretical studies. Completely new phenomena
like the coexistence of the SP state with a long - range
antiferromagnetically ordered state in slightly doped CuGeO$_3$
(Si on the Ge site or Zn on the Cu site)
\cite{Hase93a,Weiden97,Poirier95,Renard95,Regnault95,Masuda2000}
and the the strong influence of the frustration due to the next -
nearest - neighbor exchange on the magnetic and thermodynamic
properties \cite{Fabricius98} were reported. Partial substitution
of Ge by Si in CuGe$_{1-x}$Si$_{x}$O$_3$ has been an important
subject in this field, but despite considerable efforts, it was
not possible to substitute more than 50 \% Ge by Si without
changing the structure (x $\leq$ 0.1 for single crystals,
\cite{Poirier95,Renard95,Regnault95} x $\leq$ 0.5 for polycrystals
 \cite{Weiden97}). Pure CuSiO$_{3}$ was considered to be non
existent. However, recently Otto et al. \cite{Otto99} succeeded in
the synthesis of reasonable amounts of pure isostructural
CuSiO$_{3}$ by using the silicate mineral dioptase
Cu$_6$Si$_6$O$_{18}\cdot$6H$_2$O as a starting material. Compared
to the Ge homologue the unit cell volume of CuSiO$_{3}$ is reduced
($\approx$ - 3.8 \% ) due to the smaller size of Si. This
naturally leads a modification of bond angles and lengths,
\cite{Otto99} which should have a strong influence on the strength
of the magnetic interaction governed by the super - exchange
between neighboring Cu$^{2+}$ ions via the O$^{2-}$ ions. The
obvious question is how these structural changes do affect the
ground state properties of this system. Here we present the first
investigation of the physical properties of CuSiO$_{3}$, based on
susceptibility, specific heat and Cu- nuclear quadrupole resonance
experiments.

\section{Experimental details}
High quality crystals of dioptase from the type locality Altyn
Tyube (Kazakhstan) were heated up to 873 K and held at this
temperature for six hours in order to obtain dehydrated dioptase
(black dioptase). The black dioptase was decomposed by a
subsequent heat treatment at higher temperatures (1050 K) and for
20 hours under nitrogen atmosphere. The finely ground samples of
darkish brown color were characterized by X - ray powder
diffraction using the Guinier method. The diffraction pattern
reveals a mixture of three different phases. A quantitative phase
analysis reveals that about 76 wt.-\% of the mixture consist of
the new phase CuSiO$_{3}$. The other phases are CuO (tenorite,
13.7 wt.-\%) and SiO$_{2}$ (amorphous, 10.3 wt.-\%). The
orthorhombic unit - cell of the new phase CuSiO$_3$ was refined
from Guinier data giving a = 4.6357(6) \AA, b = 8.7735(11) \AA,
and c = 2.8334(4) \AA. The lattice constants of CuSiO$_3$ are in
good agreement with an extrapolation of the results from the
diluted system CuGe$_{1-x}$Si$_{x}$O$_3$ \cite{Weiden97} to x = 1.
The crystal structure was determined to be isostructural to
CuGeO$_3$. Details of the synthesis and structure characterization
are given in Ref. \onlinecite{Otto99}. Magnetization, specific
heat and NQR measurements were performed on samples taken from the
same batch. The dc - magnetization measurements at low fields
$\mu_{0}$ H $\leq$ 1 T were carried out using a commercial SQUID
magnetometer (MPMS, Quantum Design). The ac - and dc -
magnetization measurements in higher fields (1 T $\leq \mu_{0}$ H
$\leq$ 14 T) and the specific heat measurements were performed in
a commercial multi - purpose device (PPMS, Quantum Design). Ac-
and dc- magnetization are extracted from the induction signal of a
mutual inductance coil arrangement. The specific heat was
determined by standard relaxation technique and an advanced two -
tau model was applied to analyze the thermal response. The data
obtained were corrected by subtracting the specific heat
contributions of the sample holder, thermometers, heaters and
glue. The NQR measurements were performed on a conventional pulsed
spectrometer using the point - by point method. The masses of the
samples are m = 91.40 mg, 8.72 mg and 180.20 mg for
susceptibility, specific heat and NQR - measurements,
respectively.

\section{Results}

The temperature dependence of the magnetic susceptibility
$\chi$(T) of the CuSiO$_{3}$ powder sample in a magnetic field of
1 T is shown in Fig. \ref{fig:Chi}. Below room temperature,
$\chi$(T) increases with decreasing temperature, confirming
localized Cu$^{2+}$ moments. Between 200 K and 300 K the
susceptibility data follow nicely a Curie - Weiss law with a Weiss
- temperature of $\theta$ = -7.2 K, indicating rather weak
antiferromagnetic coupling. An effective moment of $\mu_{eff}$ =
1.56 $\mu_{B}$ per $Cu^{+2}$ ion is determined which corresponds
to a g - factor of 1.80. Both impurity phases CuO and $SiO_{2}$
exhibit a small susceptibility which could be neglected in a first
analysis. Scaling the measured susceptibility with the estimated
amount of pure CuSiO$_{3}$ in the sample results in an $\mu_{eff}$
- value of $1.79 \mu_{B}$ and a $g$-factor of 2.06. These values
are very close to those found for CuGeO$_{3}$ and expected
theoretically for free S = 1/2 spins with $g = 2$ and $\mu =
g\mu_{B}\sqrt{S(S+1)} = 1.73 \mu_{B}$.
\\
\\At lower temperatures the susceptibility exhibits a broad maximum
at $T_{m,\chi}$ = 13.5 K, which is a hallmark for low -
dimensional spin systems. \cite{Bonner64} Above 8 K, $\chi$(T)
could be fitted very well with the numerical results of Kl\"umper
 \cite{Klumper} for S=1/2 Heisenberg chains. The best fit is
obtained with a nearest - neighbor coupling of J/k$_{B}$ =
2$T_{m,\chi}$/1.282  \cite{Bonner64} = 21 K and a slightly
enhanced $\mu_{eff}$ - value of 1.60 $\mu_{B}$ (note that we used
the unscaled susceptibility for the fit and that we define $J$ by
the exchange Hamiltonian: $H=-J\sum \mathbf{s_{i}s_{i+1}}+$ $
\alpha\mathbf{s_{i}s_{i+2}}$ ). The quality of the fit suggests
that frustration effects, i.e. an antiferromagnetic interaction
$J'$ between next - nearest - neighbors, are negligible in
CuSiO$_{3}$ $(\alpha = J'/J = 0)$. This is in contrast to
CuGeO$_{3}$ where $\alpha$ values between 0.24 and 0.35 are
proposed. \cite{Hase93,Fabricius98} Furthermore it is evident that
the Cu - O(2) - Cu exchange in CuSiO$_{3}$ is much weaker
(J/k$_{B}$ = 21 K) than in CuGeO$_{3}$ (J/k$_{B}$ $\approx$ 160
K). Below 8 K, $\chi$(T) decreases very rapidly and saturates at
the lowest temperatures. The derivation of the susceptibility
d$\chi$/dT shows a peak at $T_{0}$ = 7.9 K which gives clear
evidence for a cooperative phase transition at this temperature
(see inset Fig.1). The signature of the transition in the
susceptibility looks more like a long - range antiferromagnetic
order in a polycrystaline sample than a spin - Peierls transition.
For the latter scenario, one expects a vanishing susceptibility at
lowest temperatures $\chi$(T $\rightarrow$ 0) = 0, which is not
observed. The absence of a Curie - like tail in the susceptibility
at the lowest temperatures indicates the absence of defects and
thus points to a high crystalline perfection of the CuSiO$_{3}$
phase.
\\
\\The presence of a phase transition at $T_{0}$ is clearly
confirmed by the specific heat results. The temperature dependence
of the total specific heat of three CuSiO$_{3}$ - crystals (total
mass of 8.72 mg, 1 mol\, =\, 139.6\, g) below 20 K is shown in
Fig. \ref{fig:specheat}. For a comparison, the data of Liu et al.
\cite{Liu95} for CuGeO$_{3}$ are plotted in the same figure. The
specific heat of CuSiO$_{3}$ shows a very clear $\lambda$ -
shaped, asymmetric anomaly at $T_{0}$ = 7.9 K with a specific heat
jump of $\triangle C\,  \approx  (1.50\,\pm \, 0.05 )\, J/molK$.
This is comparable to the $\triangle C\ $ - value of approximately
$ 1.8 J/molK$ found for CuGeO$_{3}$. \cite{Liu95} The total
specific heat obtained is the sum of the lattice term $C_{ph}(T)$
from the phonons and the magnetic term $C_{m}(T)$ from the spin
system. The separation of the two contributions is not trivial. At
low temperatures, the phonon contribution should follow a $T^{3}$
law, i.e., $C_{ph}=\beta T^{3}$. Because of the smaller mass of
the Si ions compared to the Ge ions ($m_{Si}/m_{Ge}\approx\,0.39$)
one can expect a harder phonon spectra in CuSiO$_{3}$ and thus a
reduced phonon contribution ( $\beta_{CuSiO_{3}} < \beta
_{CuGeO_{3}} \approx\, 0.32 m J/molK^{4}$ \,
\cite{Liu95,Fabricius98,Lasjaunias97}). Thus, the estimate of the
phonon contribution for CuGeO$_{3}$ (see Fig. \ref{fig:specheat},
dashed line) yields therefore an upper limit for the phonon
contribution in CuSiO$_{3}$. This clearly demonstrates, that below
20 K, the specific heat of CuSiO$_{3}$ is dominated by the
magnetic contribution. Well below $T_{0}$, C(T) follows a $T^{3}$
power law with a coefficient $\beta_{m}$ ($\approx 4.5  m
J/molK^{4}$), more than one order of magnitude larger than that
expected for the phonon contribution , indicating that it has to
be related to the magnetic interactions. Such a power law is
expected for long - range ordered 3D antiferromagnets with weak or
absent anisotropy. The $T^{3}$ power law at low temperatures is in
contrast to the experimental findings on the spin - Peierls
compound CuGeO$_{3}$ where the opening of a gap in the magnetic
excitation spectra leads to an exponential decrease of the
specific heat. Therefore our specific heat measurements, too,
support long - range antiferromagnetic order rather than a spin -
Peierls dimerization at $T_{0}$.
\\
\\At higher temperatures $T > T_{0}$, the specific heat $C_{m}(T)$ of a Heisenberg
chain without frustration exhibits a maximum at $T_{m,c} = 0.75$
$T_{m, \chi} $\cite{Bonner64} $= 10.1\, K$ with a value of
$C_{m}(T_{m,c})$ $\approx 0.35 R \approx 2.9 J/molK $, which is
independent of J. \cite{Bonner64,Klumper,Fabricius98,Johnston00}
Scaling this value with the amount of pure CuSiO$_{3}$ in the
sample results in an expected magnetic contribution of 2.2 J/molK.
The experimental value of the specific heat at 10 K is only
slightly lower. Further measurements and analysis are currently
under progress to improve the estimation of the magnetic specific
heat contribution. At even higher temperatures $(T > J/k_{B})$
$C_{m}(T)$ becomes very small and C(T) originates mainly from the
phonon contribution. Therefore the germanate exhibits a larger
total specific heat than the silicate as evidenced from
Fig.\ref{fig:specheat}.
\\
\\The influence of magnetic fields up to 14 T on the magnetic
susceptibility and the specific heat is shown in Fig.
\ref{fig:field}. In the susceptibility $ \chi (T,H)$ the signature
of the transition at $T_{0}$ = 7.9 K is smeared out, the drop in
the susceptibility is reduced and the susceptibility maximum at
$T_{m,\chi}$ is shifted significantly  to lower temperatures with
field. This shift of $T_{m,\chi}$, which indicates the suppression
of the antiferromagnetic in - chain correlations with increasing
field, is in good agreement with the theoretical calculations of
Kl\"umper \cite{Klumper} (see inset in Fig. \ref{fig:field}) for
the $S=1/2$ Heisenberg chain. In the specific heat C(T) the
transition at $T_{0}$ = 7.9 K is clearly observable in fields up
to 14 Tesla and the $T^{3}$ power law at low temperatures is
preserved. The antiferromagnetic order temperature $T_{0}$ shows
only a weak field dependence as indicated in the magnetic phase
diagram plotted in Fig. \ref{fig:PHASE}a. No other transitions are
visible in our $C(T,H)$ measurements. The temperature dependent
$\chi(T)$ measurements at fixed fields (Fig. \ref{fig:field}) and
additionally field dependent ac - susceptibilty measurements at
fixed temperatures (Fig. \ref{fig:PHASE}b) evidences a broadened
(due to the random orientation of the powder particles) transition
at $\mu_{0}H_{SF} \approx 3 T$, which looks very similar to a spin
- flop transition. The phase diagram corresponds to that observed
for the AF - phase in doped CuGeO$_{3}$, but it is quite different
from that expected and confirmed for a spin- Peierls transition.
The almost field independent transition temperature and the
presence of a spin - flop - like transition are strong evidences
for a antiferromagnetically ordered ground state in CuSiO$_{3}$.
\\
\\We have investigated the nuclear quadrupole
resonance (NQR) of Cu in CuSiO$_{3}$ to deduce microscopic
informations at nuclear sites. A spectrum at 4.2 K is shown in
Fig. \ref{fig:NQR}. The lines are fitted well by a Gaussian
function and the central frequencies $^{63,65}\nu_{NQR}$ and the
line widths $^{63,65} \triangle \nu_{NQR}$ were determined at
different temperatures (see inset of Fig.\ref{fig:NQR}). The NQR
signals at 4.2 K have been found at 26.88 $\pm$ 0.02 MHz for
$^{63}$Cu and 24.88 $\pm$ 0.02 MHz for $^{65}$Cu. The ratio of the
frequencies of the NQR signals are in good agreement with that
expected from the nuclear quadrupole moments of Cu ($^{63}Q$ /
$^{65}Q$ = 1.081 ). Also the ratio of the signal intensities
$^{63}I / ^{65}I = 2.8$ corresponds to that of the natural
abundance of the isotopes $^{63}$Cu and $^{65}$Cu with $^{63}I /
^{65}I = 2.20$. Furthermore the presence of the impurity phase CuO
is confirmed nicely by NQR measurements. At room temperature the
$^{63}$Cu NQR frequency of CuO is found at $^{63}\nu_{NQR}$ = 20.6
MHz, which is in good agreement with the literature.
\cite{Shimizu93} Due to the AF order in CuO at $T_{N}$ = 230 K
\cite{Ota92} the $^{63}\nu_{NQR}$ is shifted to much higher
frequencies below $T_{N}$ ($^{63}\nu_{NQR} \approx 137 MHz$ for
the central line at 4.2 K). \cite{Tsuda88} Therefore the observed
resonance signals presented in Fig. \ref{fig:NQR} could be clearly
assigned to the Cu NQR lines of the CuSiO$_{3}$ phase. No other
resonance lines are observed in the frequency range of 20 - 90 MHz
indicating the crystallographically equivalence of the Cu site.
Compared to CuGeO$_{3}$ with 34.23 $\pm$0.02 MHz for $^{63}Cu$ and
31.66 $\pm$0.02 MHz for $^{65}Cu$ at 4.2 K \cite{Kikuchi94,Itho95}
the NQR lines in CuSiO$_{3}$ are shifted to lower frequencies. One
possible explanation is the effect of the modified bond lengths
and angles on the electric field gradient EFG which affects
strongly $^{63,65}\nu_{NQR}$. \cite{Shimizu93} The EFG at the Cu
site originates mainly from the ionic charge distribution of the
surrounding ions (so called lattice contribution) and the 3d Cu
charge distribution (valence) itself. \cite{Shimizu93} Compared to
CuGeO$_{3}$ the Cu-O(2) bond length is nearly the same whereas the
Cu-Cu distance is reduced ( $\approx -3.5 \%$). In a simple point
charge model for the lattice contribution this could reduce the
EFG and therefore lower the NQR frequencies. Details of the NQR
results and a complex analysis of the data will be presented
elsewhere. Surprisingly $^{63,65}\nu_{NQR}$ and the line widths
$^{63,65} \triangle \nu_{NQR}$ exhibit only a weak temperature
dependence between 4.2 and 40 K and especially around the
transition at $T_{0}$ = 7.9 K, no anomaly is observed (see inset
Fig. \ref{fig:NQR}). Usually an antiferromagnetic phase transition
is associated with the appearance of strong internal magnetic
fields on the Cu site. This should result in a remarkable
transformation of the pure NQR spectrum at T $>T_{N}$ to a high
frequency AFMR spectrum for T $<T_{N}$ which is perturbed by
quadrupole interaction. The origin of this absence of any
signature of the transition is presently not clear.

\section{Discussion }

CuSiO$_3$, isotypic to the spin - Peierls compound CuGeO$_3$ was
synthesized from the mineral dioptase and the physical properties
of this new compound were determined by means of susceptibility,
specific heat and Cu nuclear quadrupole resonance measurements.
The susceptibility of CuSiO$_3$ is in good agreement with the
theoretical results for a quasi one dimensional $S = 1/2$
Heisenberg antiferromagnet with a nearest - neighbor exchange
constant of $J/k_{B}\,=\,21 K$ and without a significant next -
nearest neighbor exchange interaction ($J'\,=\,0$). This is in
contrast to the findings for CuGeO$_3$ where frustration effects
due to next - nearest neighbor interactions play a crucial role
($\alpha = J'/J \approx \, 0.35$) and where a much higher exchange
coupling constant of $J/k_{B}\,\simeq \,160 K$ was found.
According to the so - called Goodenough - Kanamori - Anderson
(GKA)\cite{Goodenough55} rules a change from an antiferromagnetic
exchange to a ferromagnetic exchange is expected when the bond
angle is near $90^{\circ}$. Therefore the much smaller J value in
CuSiO$_3$ could easily be attributed to the reduction of the Cu -
O(2) - Cu bond angle from $99 ^{\circ}$ in CuGeO$_3$ to $94
^{\circ}$ in CuSiO$_3$. In contrast, the disappearance of the
frustration is surprising, since the corresponding bond angles are
fare away from $90 ^{\circ}$. One possible explanation is, that
the smaller Cu - O(2) - Cu bond angle of $94 ^{\circ}$ leads to a
cancellation of the antiferromagnetic and ferromagnetic exchange
contributions and a vanishing nearest - neighbor exchange $J$. The
remaining next - nearest - neighbor exchange $J'$ would then
transform one atomic chain into two independent spin chains. Since
the susceptibility (and the specific heat) is the same as that of
a single spin chain, the present experimental results do not allow
to distinguish between both cases. However, theoretical
calculations do not support this scenario. \cite{Rosner2000}
\\
\\The large residual susceptibility below $T_{N}$, the $T^{3}$
power law in the magnetic specific heat at low temperatures, the
spin - flop like transition in a magnetic field as well as the H-T
magnetic phase diagram are very strong evidences for the
antiferromagnetic nature of the transition. The comparatively
large ratio between the ordering temperature $T_{N}$ and the
temperature of the maximum in the susceptibility $T_{m,\chi}$
indicate that the ratio between inter - and intra - chain exchange
( $J_{\bot}´/J$) is significantly larger in CuSiO$_3$ than in
CuGeO$_3$. This is a natural consequence of the weakness of the
intra - chain exchange. An estimate of the inter - chain coupling
constant for quasi - one - dimensional chains is given by the
following expression: \cite{Schulz96}
$J_{\bot}=\frac{T_{N}}{1.28\sqrt{ln(5.8J/T_{N})}}=3.7 K$. From
this we obtain a ratio of $J_{\bot}/J \approx 0.18$. LDA band
structure calculations also reveal a much weaker intra - chain
coupling J and a larger $J_{\bot}/J$ ratio ($ \approx
0.14$)\cite{Rosner2000} compared to CuGeO$_3$, which is in
agreement with our experimental results. However, they also
suggests that the frustration should still be significant in
CuSiO$_3$. The weaker intra - chain coupling J and the much larger
$J_{\bot}/J$ ratio are obvious responsible for the occurrence of
an AF - transition instead of a SP - transition.
\\
\\In conclusion it is shown, that CuSiO$_3$ is a  quasi - one - dimensional S=1/2 Heisenberg
chain system which undergoes a transition to long range
antiferromagnetic order at $T_{N}=7.9 K$. There is no direct
evidence for spin - Peierls transition in this new compound. Field
dependent susceptibility measurements reveal a spin - flop phase
at $\mu _{0} H_{SF} \approx 3 T$. Among all inorganic low
dimensional Cu based spin - systems the edge - sharing compound
CuSiO$_3$ exhibits the smallest intra - chain exchange AF -
coupling constant of $J/k_{B}=21 K$. The existence of the two
homologeus compounds CuGeO$_3$ and CuSiO$_3$ with different ground
state properties provides an excellent basis for the application
of theoretical models and methods of low dimensional physics.

\section*{Acknowledgments}
We acknowledge fruitful discussions with H. Rosner and S.-L.
Drechsler.

\begin{figure}
\caption{Raw data of the magnetic susceptibility of a CuSiO$_3$
sample measured under a field of $\mu_0$H = 1 T in a log-T scale.
The solid curve corresponds to the numerical calculations of
Kl\"umper et al. [4] for S = 1/2 1D - Heisenberg chains with
nearest - neighbor coupling and without next - nearest - neighbor
coupling (J' = 0). The fitting of the numerical results to the
experimental $\chi$(T) - data requires a coupling constant J/k$_B$
= 21 K and an effective moment of $\mu_{eff} = 1.60 \mu_B$ ($g =
1.85$). At high temperatures T $\geq$ 200 K, $\chi$(T) is fitted
with a Curie - Weiss law (dashed line) with $\mu_{eff}=1.56\mu_B$
(g = 1.80) and $\theta$ = -7.2 K. The inset shows ${\rm
d}\chi/{\rm d}T$ as a function of temperature.} \label{fig:Chi}
\end{figure}

\begin{figure}
\caption{Heat capacity of a CuSiO$_{3}$ sample and of CuGeO$_{3}$
(data taken from  Ref. [13]. The dashed line represents the phonon
contribution $C_{ph}=\beta T^3$ ($\beta =0.32 m J/mol K^4$) for
CuGeO$_{3}$.} \label{fig:specheat}
\end{figure}

\begin{figure}
\caption{Temperature dependence of the specific heat (left side,
note that the origin of the vertical axis is shifted by 1.6 J/molK
for the curves with $\mu_{0}H  \geq 0 T$) and the magnetic
susceptibility (right) for different magnetic fields. The inset
shows the shift of the temperature $T_{m,\chi}$ defined by the
maximum susceptibility $\chi (T_{m,\chi}) = \chi_{max}$ in reduced
units as a function of applied field and in comparison with the
theoretical prediction of Kl\"umper et al. .[4] }
\label{fig:field}
\end{figure}

\begin{figure}
\caption{ Magnetic phase diagram of CuSiO$_{3}$ probed by specific
heat and susceptibility measurements (left). $H_{c}$ denotes the
AF transition whereas $H_{SF}$ indicates the spin flop transition
as evidenced from ac - susceptibility measurements (left side,
note that the origin of the vertical axis is shifted by a constant
value for curves at different temperatures).} \label{fig:PHASE}
\end{figure}

\begin{figure}
\caption{  $^{63}$ Cu and $^{65}$ Cu NQR spectra in CuSiO$_{3}$ at
4.2 K. The solid curves corresponds to the fitting of a Gaussian
function to the experimental data. The inset shows the temperature
dependence of the $^{63}$ Cu NQR frequency (right axis) and of the
$^{63}$ Cu NQR line widths (left axis).} \label{fig:NQR}
\end{figure}

\end{document}